\documentclass[conference]{IEEEtran}
\IEEEoverridecommandlockouts
\usepackage{cite}
\usepackage{amsmath,amssymb,amsfonts,color}
\usepackage{algorithm,algpseudocode}
\usepackage{algorithmicx}
\usepackage{graphicx}
\usepackage{textcomp}
\usepackage{xcolor,soul}
\usepackage{tikz}
\usepackage{stfloats}
\usepackage{url}
\usepackage{booktabs}
\usepackage[normalem]{ulem}
\useunder{\uline}{\ul}{}
\usepackage{array,ragged2e}
\usepackage[bookmarks=false]{hyperref}
\usepackage{trimclip} 
\usepackage{xcolor}

\ifCLASSOPTIONcompsoc
    \usepackage[caption=false, font=normalsize, labelfont=sf, textfont=sf, subrefformat=parens, labelformat=parens]{subfig}
\else
\usepackage[caption=false, font=footnotesize, subrefformat=parens, labelformat=parens]{subfig}
\fi

\def\BibTeX{{\rm B\kern-.05em{\sc i\kern-.025em b}\kern-.08em
    T\kern-.1667em\lower.7ex\hbox{E}\kern-.125emX}}

\usepackage[nodisplayskipstretch]{setspace}

\algnewcommand{\IfThenElse}[3]{
  \State \algorithmicif\ #1\ \algorithmicthen\ #2\ \algorithmicelse\ #3}

\algnewcommand{\IfThen}[2]{
  \State \algorithmicif\ #1\ \algorithmicthen\ #2}  

\usepackage{mathtools} 
\usepackage{xspace}

\newcommand{\Secref}[1]{Section~\ref{#1}}

\newcommand{\ba}{\mathbf{a}}

\newcommand{\bh}{\mathbf{h}}

\newcommand{\bp}{\mathbf{p}}

\newcommand{\br}{\mathbf{r}}

\newcommand{\bw}{\mathbf{w}}

\newcommand{\bz}{\mathbf{z}}

\newcommand{\bG}{\mathbf{G}}
\newcommand{\bH}{\mathbf{H}}
\newcommand{\bI}{\mathbf{I}}

\newcommand{\bP}{\mathbf{P}}

\newcommand{\bR}{\mathbf{R}}

\newcommand{\bW}{\mathbf{W}}

\newcommand{\bOmega}{\boldsymbol{\Omega}}

\newcommand{\bEtad}{\boldsymbol{\eta}}

\newcommand{\bzero}{\boldsymbol{0}}

\newcommand{\re}[1]{\mathfrak{R}\left\{{#1}\right\}}

\newcommand{\diag}{\text{diag}}

\newcommand{\herm}{^{\mathsf{H}}}
\newcommand{\trans}{^\mathsf{T}}

\DeclareMathOperator{\tr}{tr}
\DeclareMathOperator{\E}{\mathsf{E}}

\newcommand{\EX}[1]{\E\left\{{#1}\right\}}

\newcommand*{\imagj}{\mathsf{j}} 
\newcommand*{\myexp}{\mathsf{e}}
\newcommand{\norm}[1]{{ \left\Vert #1 \right\Vert }}

\newcommand{\mC}{\mathbb{C}}
\newcommand{\mR}{\mathbb{R}}

\newcommand{\CG}[2]{\mathcal{CN}\left({#1},{#2}\right)}

\newcommand{\etad}{\eta}

\newcommand{\Na}{N_{\text{A}}}
\newcommand{\Nr}{N_{\text{R}}}

\newcommand{\RIS}{\text{RIS}}

\newcommand{\tR}{{\text{R}}}



\usepackage{graphicx}
\graphicspath{{fig/}}
\usepackage{textcomp}  
\usepackage{listings}  
\usepackage{tabulary}
\usepackage{array}
\usepackage{booktabs}

\makeatletter
\g@addto@macro\normalsize{%
  \setlength\abovedisplayskip{3.2pt}
  \setlength\belowdisplayskip{3.2pt}
}

\begin{document}
	
\bstctlcite{IEEE_nodash:BSTcontrol}
\bstctlcite{IEEEexample:BSTcontrol}

\title{Evaluating the Performance of Reconfigurable Intelligent Base Stations through Ray Tracing 
}
\author{\IEEEauthorblockN{Sina Beyraghi\IEEEauthorrefmark{1}\IEEEauthorrefmark{2}, Giovanni Interdonato\IEEEauthorrefmark{3}, Giovanni Geraci\IEEEauthorrefmark{1}\IEEEauthorrefmark{2}, Stefano Buzzi\IEEEauthorrefmark{3}\IEEEauthorrefmark{4}, Angel Lozano\IEEEauthorrefmark{2}}
\IEEEauthorblockA{\IEEEauthorrefmark{1}\textit{Telefonica Scientific Research, Barcelona, Spain}}
\IEEEauthorblockA{\IEEEauthorrefmark{2}\textit{Dept. of Engineering, Universitat Pompeu Fabra, Barcelona, Spain}}
\IEEEauthorblockA{\IEEEauthorrefmark{3}\textit{Dept. of Electrical and Information Engineering, University of Cassino and Southern Lazio, 03043 Cassino, Italy.}\\
\IEEEauthorrefmark{4}\textit{Dept. of Electronics, Information and Bioengineering, Politecnico di Milano, 20133 Milan, Italy.}\\
\texttt{\small \{mohammadsina.beyraghi,giovanni.geraci\}@telefonica.com}}
\texttt{\small\{giovanni.interdonato,buzzi\}@unicas.it}, 
\texttt{\small angel.lozano@upf.edu}
\thanks{This work was supported by the EU Horizon 2020 MSCA-ITN-METAWIRELESS (GA 956256), and the SNS (Smart Networks and Services) JU and its members, funded by the European Union under Grant Agreement number 101139161.}}%

\maketitle

\begin{abstract}
Massive multiple-input multiple-output (mMIMO) is a key capacity-boosting technology in 5G wireless systems. To reduce the number of radio frequency (RF) chains needed in such systems, a novel approach has recently been introduced involving an antenna array supported by a reconfigurable intelligent surface. This arrangement, known as a reconfigurable intelligent base station (RIBS), offers performance comparable to that of a traditional mMIMO array, but with significantly fewer RF chains. 
Given the growing importance of precise, location-specific performance prediction, this paper evaluates the performance of an RIBS system by means of the SIONNA ray-tracing module. That performance is contrasted against results derived from a statistical 3GPP-compliant channel model, optimizing power and RIS configuration to maximize the sum spectral efficiency. Ray tracing predicts better performance than the statistical model in the evaluated scenario, suggesting the potential of site-specific modeling. However, empirical validation is needed to confirm this advantage.
\end{abstract}

\begin{IEEEkeywords}
Reconfigurable intelligent surface, RIS, massive MIMO, ray tracing, near-field communications, optimization.
\end{IEEEkeywords}

\section{Introduction}

Massive multiple-input multiple-output (mMIMO) is crucial for 5G networks~\cite{redbook}. It relies on numerous active antennas and extensive digital baseband processing at the base station (BS), to spatially multiplex user equipment (UE) units within the same time-frequency resource, enhancing coverage, spectral, and energy efficiency.
However, excessively increasing the number of active antennas is costly and energy inefficient, as energy consumption rises linearly with the number of radio frequency (RF) chains.
Reconfigurable intelligent surfaces (RISs) are an affordable solution for enhancing performance and overcoming challenging propagation conditions~\cite{Huang2019}.An RIS uses low-cost, passive elements, reconfigured in real time, to steer energy and prevent interference. Initially, RIS technology only allowed for phase-shift tuning in reflected waves to boost single and multiuser MIMO systems~\cite{Yan_JSAC2020}. Later, active RISs were developed to control both the amplitude and phase of the reflected waves~\cite{Long2021,ZhangZ2023}.

In \cite{interdonato2024RISaided}, the authors investigated a novel setup involving a BS with a planar antenna array that integrates a RIS to support multiple UEs, as illustrated in Fig. 1. This setup is termed a \textit{reconfigurable intelligent base station} (RIBS). Their findings indicate that RIBS facilitates the multiplexing efficiencies and performance of standard MIMO systems using far fewer active antennas and RF chains. The research particularly highlighted the significance of employing active RISs, which markedly surpass, for the same amount of total radiated power, the performance of RIBS based on passive RISs.

Recently, there has been growing interest in using ray tracing to generate accurate site-specific electromagnetic models\cite{heng2024site}. This study concentrates on assessing the performance achievable via a RIBS when using the SIONNA ray tracing module \cite{hoydis2023sionna} to characterize electromagnetic propagation within a specified region. We compare the resultant performance with that derived from a traditional 3GPP-compliant channel model when power and RIS configuration are optimized to maximize the system sum spectral efficiency (SE). Notably, our findings show that ray tracing predicts higher performance than statistical models in the simulated scenario, which may suggest that transceiver algorithms tailored to specific sites could help improve resource utilization. However, these observations require empirical validation.


\section{System model}

\begin{figure}
    \centering
       \resizebox{.8\columnwidth}{!}{
  
\tikzset {_gbrv52ujj/.code = {\pgfsetadditionalshadetransform{ \pgftransformshift{\pgfpoint{0 bp } { 0 bp }  }  \pgftransformrotate{-23 }  \pgftransformscale{2 }  }}}
\pgfdeclarehorizontalshading{_mx4rqpzab}{150bp}{rgb(0bp)=(1,0,0);
rgb(37.5bp)=(1,0,0);
rgb(43.75bp)=(1,1,0);
rgb(50bp)=(0.02,0.76,1);
rgb(56.25bp)=(1,1,0);
rgb(62.5bp)=(1,0,0);
rgb(100bp)=(1,0,0)}
\tikzset{every picture/.style={line width=0.75pt}} 

\begin{tikzpicture}[x=0.75pt,y=0.75pt,yscale=-1,xscale=1]

\draw  [fill={rgb, 255:red, 128; green, 128; blue, 128 }  ,fill opacity=1 ] (205.42,69.87) -- (205.42,166.48) .. controls (205.42,167.19) and (203.51,167.76) .. (201.16,167.76) .. controls (198.8,167.76) and (196.89,167.19) .. (196.89,166.48) -- (196.89,69.87) .. controls (196.89,69.16) and (198.8,68.59) .. (201.16,68.59) .. controls (203.51,68.59) and (205.42,69.16) .. (205.42,69.87) .. controls (205.42,70.58) and (203.51,71.15) .. (201.16,71.15) .. controls (198.8,71.15) and (196.89,70.58) .. (196.89,69.87) ;
\draw  [fill={rgb, 255:red, 128; green, 128; blue, 128 }  ,fill opacity=1 ] (253.26,77.01) -- (248.94,73.63) -- (158.2,101.24) -- (158.63,164.61) -- (162.95,167.98) -- (253.69,140.38) -- cycle ; \draw   (158.2,101.24) -- (162.52,104.61) -- (253.26,77.01) ; \draw   (162.52,104.61) -- (162.95,167.98) ;
\path  [shading=_mx4rqpzab,_gbrv52ujj] (253.48,140.44) -- (253.14,77.09) -- (162.89,104.73) -- (163.23,168.08) -- cycle ; 
 \draw   (253.48,140.44) -- (253.14,77.09) -- (162.89,104.73) -- (163.23,168.08) -- cycle ; 

\draw [color={rgb, 255:red, 155; green, 155; blue, 155 }  ,draw opacity=1 ]   (253.44,85.01) -- (162.41,112.38) ;
\draw [color={rgb, 255:red, 155; green, 155; blue, 155 }  ,draw opacity=1 ]   (252.75,93.63) -- (163.23,120.23) ;
\draw [color={rgb, 255:red, 155; green, 155; blue, 155 }  ,draw opacity=1 ]   (252.8,101.59) -- (163.29,128.2) ;
\draw [color={rgb, 255:red, 155; green, 155; blue, 155 }  ,draw opacity=1 ]   (252.74,110.53) -- (163.22,137.14) ;
\draw [color={rgb, 255:red, 155; green, 155; blue, 155 }  ,draw opacity=1 ]   (252.87,118.35) -- (163.35,144.95) ;
\draw [color={rgb, 255:red, 155; green, 155; blue, 155 }  ,draw opacity=1 ]   (252.87,126.87) -- (163.35,153.48) ;
\draw [color={rgb, 255:red, 155; green, 155; blue, 155 }  ,draw opacity=1 ]   (252.87,134.56) -- (163.35,161.17) ;
\draw [color={rgb, 255:red, 155; green, 155; blue, 155 }  ,draw opacity=1 ]   (170.5,102.47) -- (170.92,165.86) ;
\draw [color={rgb, 255:red, 155; green, 155; blue, 155 }  ,draw opacity=1 ]   (178.82,100.39) -- (179.23,163.79) ;
\draw [color={rgb, 255:red, 155; green, 155; blue, 155 }  ,draw opacity=1 ]   (187.34,97.69) -- (187.76,161.08) ;
\draw [color={rgb, 255:red, 155; green, 155; blue, 155 }  ,draw opacity=1 ]   (194.82,95.19) -- (195.24,158.59) ;
\draw [color={rgb, 255:red, 155; green, 155; blue, 155 }  ,draw opacity=1 ]   (203.55,92.08) -- (203.97,155.47) ;
\draw [color={rgb, 255:red, 155; green, 155; blue, 155 }  ,draw opacity=1 ]   (212.07,89.58) -- (212.49,152.98) ;
\draw [color={rgb, 255:red, 155; green, 155; blue, 155 }  ,draw opacity=1 ]   (220.39,87.35) -- (220.8,150.75) ;
\draw [color={rgb, 255:red, 155; green, 155; blue, 155 }  ,draw opacity=1 ]   (228.5,84.44) -- (228.91,147.84) ;
\draw [color={rgb, 255:red, 155; green, 155; blue, 155 }  ,draw opacity=1 ]   (237.02,81.74) -- (237.43,145.14) ;
\draw [color={rgb, 255:red, 155; green, 155; blue, 155 }  ,draw opacity=1 ]   (245.33,79.43) -- (245.75,142.82) ;
\draw  [fill={rgb, 255:red, 128; green, 128; blue, 128 }  ,fill opacity=1 ] (289.83,83.14) -- (286,81.16) -- (227.36,99.78) -- (235.49,125.36) -- (239.31,127.34) -- (297.95,108.72) -- cycle ; \draw   (227.36,99.78) -- (231.19,101.76) -- (289.83,83.14) ; \draw   (231.19,101.76) -- (239.31,127.34) ;
\draw  [fill={rgb, 255:red, 128; green, 128; blue, 128 }  ,fill opacity=1 ] (196.89,70.06) -- (201.79,71.53) -- (206.7,70.06) -- (271.94,88.54) -- (248.66,96.02) -- cycle ;
\draw  (272,256.59) -- (446.38,256.59)(322.38,69) -- (322.38,269.47) (439.38,251.59) -- (446.38,256.59) -- (439.38,261.59) (317.38,76) -- (322.38,69) -- (327.38,76)  ;
\draw  [fill={rgb, 255:red, 128; green, 128; blue, 128 }  ,fill opacity=1 ] (319.8,120) -- (324.8,120) -- (324.8,189.47) -- (319.8,189.47) -- cycle ;
\draw [color={rgb, 255:red, 155; green, 155; blue, 155 }  ,draw opacity=1 ] [dash pattern={on 0.84pt off 2.51pt}]  (324.8,120) -- (381.27,120) ;
\draw [color={rgb, 255:red, 155; green, 155; blue, 155 }  ,draw opacity=1 ] [dash pattern={on 0.84pt off 2.51pt}]  (388.27,87.88) -- (388.27,126.67) ;
\draw  [fill={rgb, 255:red, 128; green, 128; blue, 128 }  ,fill opacity=1 ] (372.3,108.3) -- (376.1,105.64) -- (393.69,130.67) -- (389.89,133.34) -- cycle ;
\draw   (409.47,219.57) .. controls (409.47,218.46) and (410.36,217.57) .. (411.47,217.57) .. controls (412.57,217.57) and (413.47,218.46) .. (413.47,219.57) .. controls (413.47,220.67) and (412.57,221.57) .. (411.47,221.57) .. controls (410.36,221.57) and (409.47,220.67) .. (409.47,219.57) -- cycle ;
\draw   (415.57,198.57) .. controls (417.17,198.57) and (418.47,199.87) .. (418.47,201.47) -- (418.47,221.67) .. controls (418.47,223.27) and (417.17,224.57) .. (415.57,224.57) -- (406.87,224.57) .. controls (405.27,224.57) and (403.97,223.27) .. (403.97,221.67) -- (403.97,201.47) .. controls (403.97,199.87) and (405.27,198.57) .. (406.87,198.57) -- cycle ;
\draw  [dash pattern={on 4.5pt off 4.5pt}]  (332.68,157.28) .. controls (371.88,182.08) and (372.17,174.19) .. (399.38,196.59) ;
\draw [line width=1.5]  [dash pattern={on 5.63pt off 4.5pt}]  (332,146.31) .. controls (349,135.48) and (361,129.48) .. (376.48,123.68) ;
\draw  [draw opacity=0] (380.52,111.74) .. controls (380.81,109.76) and (381.76,108.14) .. (383.4,107.18) .. controls (384.82,106.34) and (386.59,106.09) .. (388.5,106.36) -- (392.79,118.98) -- cycle ; \draw   (380.52,111.74) .. controls (380.81,109.76) and (381.76,108.14) .. (383.4,107.18) .. controls (384.82,106.34) and (386.59,106.09) .. (388.5,106.36) ;  
\draw [color={rgb, 255:red, 155; green, 155; blue, 155 }  ,draw opacity=1 ] [dash pattern={on 4.5pt off 4.5pt}]  (388.27,83.88) -- (321.2,83.88) ;
\draw [shift={(388.27,83.88)}, rotate = 360] [color={rgb, 255:red, 155; green, 155; blue, 155 }  ,draw opacity=1 ][line width=0.75]    (0,5.59) -- (0,-5.59)   ;
\draw   (317,256.47) .. controls (317,253.82) and (319.15,251.67) .. (321.8,251.67) .. controls (324.45,251.67) and (326.6,253.82) .. (326.6,256.47) .. controls (326.6,259.13) and (324.45,261.27) .. (321.8,261.27) .. controls (319.15,261.27) and (317,259.13) .. (317,256.47) -- cycle ;
\draw  [dash pattern={on 0.84pt off 2.51pt}]  (163.23,168.08) -- (319.8,189.47) ;
\draw  [dash pattern={on 0.84pt off 2.51pt}]  (162.52,104.61) -- (319.8,120) ;

\draw (381,229) node [anchor=north west][inner sep=0.75pt]  [font=\normalsize] [align=left] {$\displaystyle k$-th UE};
\draw (375,85) node [anchor=north west][inner sep=0.75pt]   [align=left] {$\displaystyle \alpha $};
\draw (357,134) node [anchor=north west][inner sep=0.75pt]   [align=left] {$\displaystyle \mathbf{H}$};
\draw (385,165) node [anchor=north west][inner sep=0.75pt]   [align=left] {$\displaystyle \mathbf{h}_{k}$};
\draw (447,260) node [anchor=north west][inner sep=0.75pt]   [align=left] {$\displaystyle z$};
\draw (302,57) node [anchor=north west][inner sep=0.75pt]   [align=left] {$\displaystyle y$};
\draw (351,64) node [anchor=north west][inner sep=0.75pt]  [color={rgb, 255:red, 128; green, 128; blue, 128 }  ,opacity=1 ] [align=left] {$\displaystyle D$};
\draw (328,175) node [anchor=north west][inner sep=0.75pt]   [align=left] {$\displaystyle \mathbf{P}$};
\draw    (180,210) -- (292.8,210) -- (292.8,246.6) -- (180,246.6) -- cycle  ;
\draw (289.8,242.6) node [anchor=south east] [inner sep=0.75pt]  [font=\normalsize] [align=left] {\begin{minipage}[lt]{76.71pt}\setlength\topsep{0pt}
\begin{center}
{\small cascaded channel}\\$\bar{\mathbf{h}}_{k} = \mathbf{HP} \mathbf{h}_{k}$
\end{center}

\end{minipage}};
\draw (306,265) node [anchor=north west][inner sep=0.75pt]   [align=left] {$\displaystyle x$};

\end{tikzpicture}}
  \caption{Illustration of the RIBS from \cite{interdonato2024RISaided}: a planar array is mounted at a distance $D$ from a RIS, with a tilt angle $\alpha$ relative to the RIS plane.}
  \label{fig:RIBS}
  \vspace{-4mm}
\end{figure}
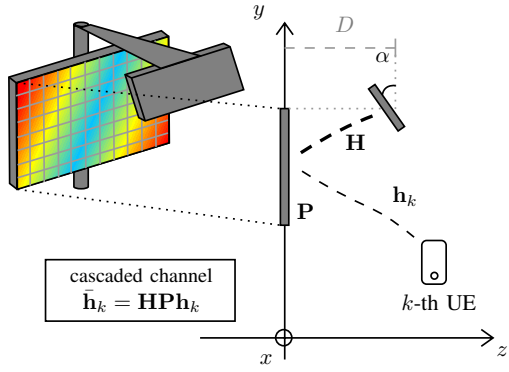

Consider a single-cell network, operating in time division duplexing at sub-6 GHz frequencies, where $K$ single-antenna UEs are served by a RIBS, as depicted in Fig.~\ref{fig:RIBS}. 

We denote by $\Na$ the number of elements of the planar array and by $\Nr$ the number of reconfigurable reflective elements at the RIS, which we assume to be active. The BS array is at a distance $D$ from the RIS and with a tilt angle $\alpha$ relative to the RIS plane. 
The effects of the active RIS on the channel are modeled as a diagonal $(\Nr \times \Nr)$ matrix, denoted by $\bP$, whose diagonal entries have a tunable magnitude that satisfies a budget power constraint. The RIS is controlled by the BS, which can optimize its configuration as required.

We consider a narrowband block-fading channel, as detailed in the remainder of this section.

\subsection{BS-to-RIS Channel}

The channel between the BS planar array and the RIS, labeled $\bH \in \mC^{\Na \times \Nr}$, is based on these assumptions: $(i)$ the BS does not block the reradiation from the RIS; $(ii)$ the direct links from the BS to the UEs are negligible relative to the cascaded links through the RIS; $(iii)$
given the short distance between the RIS and the BS and the non-varying propagation conditions, the channel is deterministic and known by the BS; and $(iv)$ the spacings between antennas and elements in the planar array and in the RIS, respectively, are at least a half-wavelength, so there is no mutual coupling. 
    
The channel between the planar array and the RIS is a pure line-of-sight (LoS) near-field channel. The $(m,n)$th entry of $\bH$ is given by 
\begin{align}
[\bH]_{m,n} &= \frac{\lambda}{4\pi d_{m,n}} \, e^{-\imagj 2\pi d_{m,n}/\lambda},
\label{eq:Hnear}
\end{align}
where $\lambda$ is the wavelength, and $d_{m,n}$ is the distance between the $m$th BS antenna and the $n$th RIS element.\footnote{A more accurate model for the channel amplitude in~\eqref{eq:Hnear} is provided in~\cite[Appendix A]{Bjornson2020NearField} and specialized for the RIBS architecture in~\cite{interdonato2024RISaided}.}

\subsection{RIS-to-UE Channel}
The channel connecting the RIS with UE $k$ is embodied by an $\Nr$-dimensional vector denoted by $\bh_k$. In the following, we present the 3GPP-compliant statistical model and the one based on the SIONNA ray tracing module. 

\subsubsection{Spatially Correlated Rician Fading}
\label{subsec:statistical-channel-model}

Spatially correlated fading is considered, with an arbitrary number $S_k$ of deterministic components~\cite{Demir2021}, that is
\begin{align}
    \bh_k = \sum^{S_k}_{s=1} \myexp^{\imagj \theta_{k,s}} \hat{\bh}_{k,s} + \tilde{\bh}_k\ \; \in \mC^{\Nr \times 1},
    \label{eq:channel-model}
\end{align}
where 
\begin{equation} \label{eq:channel_barh_k}
    \hat{\bh}_{k,s} = \sqrt{\beta_k \frac{\kappa_k}{\kappa_k+1}} \ba(\varphi_{k,s},\vartheta_{k,s}),
\end{equation}
with $\ba(\varphi_{k,s},\vartheta_{k,s})$ being the response vector of the planar RIS array, which depends on the azimuth, $\varphi_{k,s}$, and elevation $\vartheta_{k,s}$ angles-of-arrival (AoA). Moreover, $\beta_k$ is the large-scale fading channel gain that encompasses pathloss and shadowing, and $\kappa_k$ is the Rician factor.
The non-LoS (NLoS) components follow a correlated Rayleigh fading distribution and are captured by the vector $\tilde{\bh}_k \sim \CG{\bzero}{\frac{\beta_k}{\kappa_k+1}\bR_k}$, where $\bR_k \in \mC^{\Nr\times\Nr}$ is the spatial correlation matrix normalized such that $ \tr(\bR_k) = \Nr$ and modeled according to the \textit{local scattering model}~\cite[Sec. VIII]{Demir2021}. (Details are omitted for the sake of brevity.)

If there is a LOS path between the RIS and UE $k$, then one of the deterministic components $\myexp^{\imagj \theta_{k,s}} \hat{\bh}_{k,s}$ corresponds to this path. In turn, $\theta_{k,s} \sim \mathcal{U}[0,2\pi)$ is the independent random phase shift induced by the UE mobility and affecting the $s$th deterministic component. 
$\{\hat{\bh}_{k,s}\}$ and $\{\bR_k\}$ are taken to be constant on a much longer timescale than  $\tilde{\bh}_k$, which changes across channel coherence blocks. 
All the channel statistics are assumed to be known at the RIBS. 

\subsubsection{SIONNA Ray Tracing Module}
\label{subsec:ray-tracing-channel-model}

\begin{figure}[t]
    \centering
    \includegraphics[width=0.67\columnwidth]{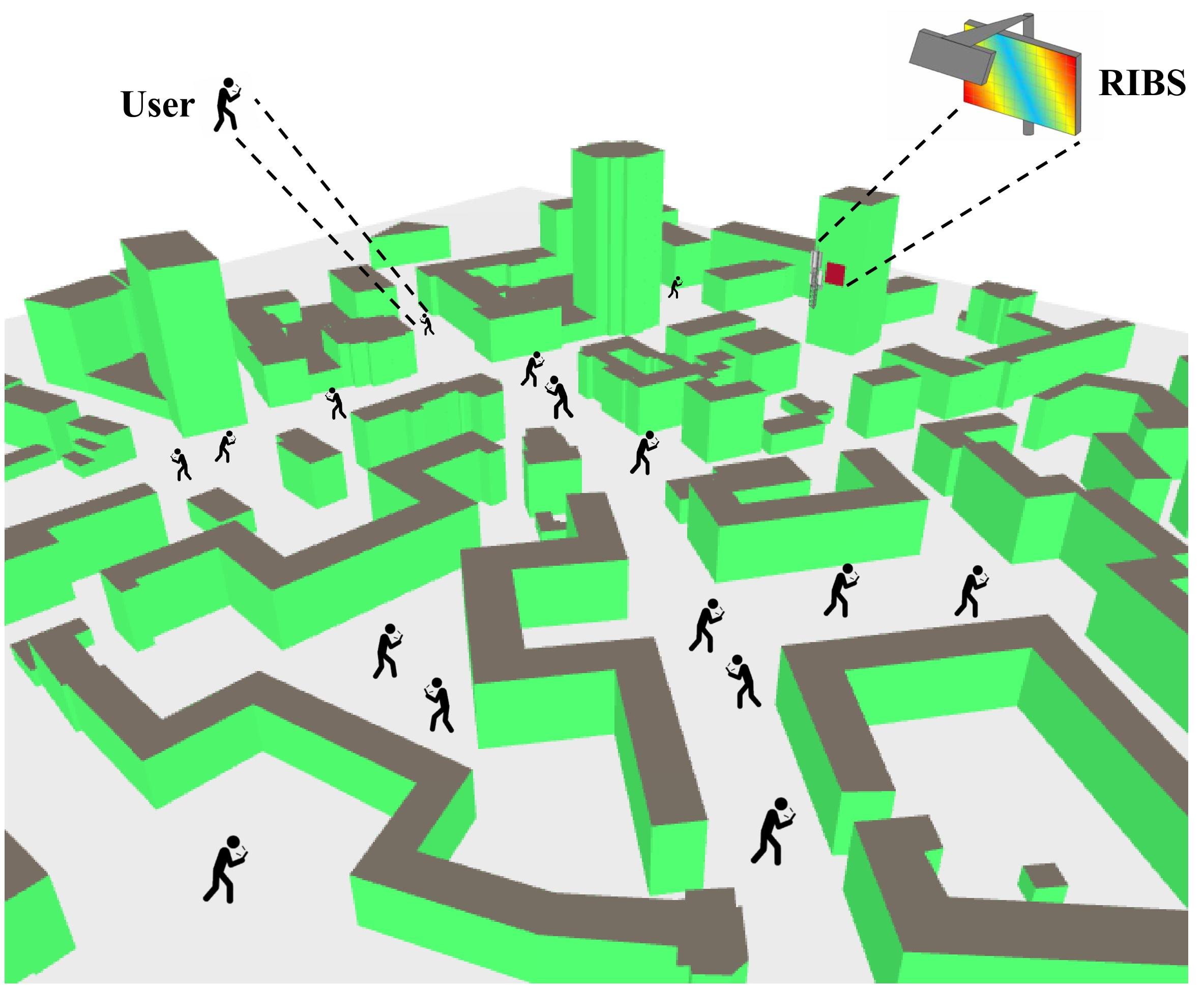}
    \caption{3D visualization of the ray-tracing simulation layout.}
    \label{ray_tracing}
    \vspace{-1.5mm}
\end{figure}
The Sionna ray tracing framework is a GPU-accelerated, open-source library based on TensorFlow, able to compute the far-field channels between RIS and UEs. Our simulations were conducted in a detailed 3D model of a selected area in London, between latitudes \([51.4969,\, 51.4931]\) and longitudes \([-0.0983,\, -0.0916]\). Fig.~\ref{ray_tracing} presents the 3D model of this area, highlighting a deployed BS, a RIS, and several randomly distributed outdoor UEs. The imported layout includes buildings and terrain, with their constituent materials set to \texttt{itu\_concrete} and \texttt{itu\_very\_dry\_ground}, respectively. These material models comply with the ITU-R recommendation P.2040-1~\cite{series2015effects}.\!

To compute the channels between the RIS and UEs in the SIONNA environment, we replaced the RIS with a transmitter equipped with a planar array that has the same number of elements as the RIS configuration. Each element is configured with an isotropic pattern to emulate passive reflection. 

Following the deployment of the transmitter and outdoor UEs, ray tracing is executed using the parameters summarized in Table~\ref{ray tracing parameters}.  In order to generate the ray-tracing-based channel vector from the RIS to UE $k$, we generate the channel from each element of the virtual planar array (that plays RIS's role) to the antenna of UE $k$.
Each ray $r$ departing from the $\ell$-th element of the RIS and arriving at UE $k$ is characterized by the complex gain \( a_r^{(\ell,k)} \) and a delay \( \tau_r^{(\ell,k)} \).
Then, based on the set of rays \( R^{(\ell,k)} \) from the $\ell$-th RIS element to UE \( k \), the baseband equivalent channel impulse response is:
\begin{equation}
h^{(\ell,k)}(\tau) = \sum_{r=1}^{R^{(\ell,k)}} a_r^{(\ell,k)}  \delta(\tau - \tau_r^{(\ell,k)}).
\label{eq:impulse_response}
\end{equation}
The channel frequency response \( H^{(\ell,k)}(f) \) is thus obtained as 
\begin{equation}
H^{(\ell,k)}(f) = \sum_{r=1}^{R^{(\ell,k)}} a_r^{(\ell,k)} \, e^{-j 2 \pi f \tau_r^{(\ell,k)}}.
\label{eq:frequency_response}
\end{equation}
By evaluating \eqref{eq:frequency_response} at the desired sub-carrier frequency, we obtain, varying $\ell=1, \ldots, N_R$, the SIONNA-generated elements of the vector $\bh_k$. 
\begin{table}[t]
\small
\centering
\caption{Ray Tracing Parameters}
\label{ray tracing parameters}
\def\arraystretch{1.2}
\begin{tabular}{ |m{.31\linewidth}|m{.59\linewidth}|}
\hline
\textbf{Ray\! tracing\! method}       & Fibonacci, shoot-and-bounce approach \\ \hline
\textbf{Activated ray types}      & LOS, specular-reflection, diffraction \\ \hline
\textbf{Number of rays}  & $16 \times 10^6$ \\ \hline
\textbf{Max\! bounces\! per\! ray}   & 4, covering both specular-reflection and diffraction \\ \hline
\textbf{Carrier frequency}    & 1.9 GHz \\ \hline
\end{tabular}
\vspace{-4mm}
\end{table}




\subsection{BS-to-UE Cascade Channel}

The cascaded uplink (UL) channel from UE $k$ to the BS, through the RIS, is the $\Na$-dimensional vector $\bar{\bh}_k = \bH \bP \bh_k$.
By letting $\bp \!=\! \diag(\bP)\!=\![p_1,\ldots,p_{\Nr}]\trans$, $\bH_k \!=\!\diag(\bh_k) \!\in\!\mC^{\Nr\times\Nr}$, and  $\bG_k \!\triangleq\! \bH \bH_k \!\in\! \mC^{\Na \times \Nr}$, note that we also have 
\begin{equation}
\bar{\bh}_k  = \bH \bH_k \bp = \bG_k \bp\,.
\end{equation}

\section{Performance Measures and Optimization}
\label{sec:performance-formulation}

The downlink (DL) signal-to-interference-plus-noise ratio (SINR) at UE $k$ under the assumption of perfect channel knowledge at the RIBS and at the UEs is given by
\begin{equation}
	\gamma_{k} (\bp) \!=\! \frac{\etad_k \left| \bp\trans \bG\trans_k \bw_k\right|^2}{\sum\nolimits_{j\neq k}^K \eta_j\left|\bp\trans	\bG_k\trans \bw_j \right|^2 \!+\! \norm{\bp\trans \bH_k}^2\! \sigma^2_\tR \!+\! \sigma^2_k} \, ,
	\label{eq:DL_SINR_PCSI}
\end{equation}
where $\etad_k$ is the transmit power reserved for UE $k$, $\bw_k \!\in\!\mC^{\Na}$ is the precoder, $\norm{\bw_k}\!=\!1$, $\sigma^2_\tR$ denotes the variance of the dynamic noise introduced by the active RIS, and $\sigma^2_k$ represents the variance of the white thermal noise at the $k$th UE.

If the channel model is stochastic, as per~\Secref{subsec:statistical-channel-model}, the achievable ergodic SE for UE $k$, in bit/s/Hz, is
\begin{equation}
	\text{SE}_{k} = \EX{\log_2 \big(1 + \gamma_{k}(\bp)\big)},
	\label{eq:DL_SE_PCSI}
\end{equation}
where the expectation is with respect to the channel realizations.
If the channel model is deterministic, as per~\Secref{subsec:ray-tracing-channel-model}, then the SE is given by~\eqref{eq:DL_SE_PCSI} but without expectation.

\subsection{Optimization of the RIS Configuration}
\label{subsec:RIS_optimization}

Thanks to the RIBS, the propagation environment can be conveniently shaped by letting the BS optimize the RIS phase shifts and amplification factors. In this paper, we formulate a DL sum-rate
maximization problem for a RIBS-aided MIMO system. By exploiting fractional programming (FP), a closed-form expression for the optimal RIS configuration is provided along with a heuristic power allocation strategy.

Due to the use of active components, the active RIS consumes additional power to amplify the reflected signals. This reflect power, denoted by $P_{\RIS}$, is a function of $\bp$ and of the BS transmit power, i.e., $\bEtad \!=\! [\etad_1,\ldots,\etad_K]\trans$.
The signal, coming from the BS, reflected by the active RIS is given by 
\begin{equation}
\br_\RIS \!=\! \sum\nolimits_{j=1}^K \sqrt{\etad_j} \bP \bH\trans \bw_j x_j \!+\! \bP \bz_\tR \, ,
\end{equation}
where $\bP \bz_\tR$, $\bz_\tR\!\sim\!\CG{\bzero_{\Nr}}{\sigma^2_\tR\bI_{\Nr}}$, denotes the dynamic noise introduced by the active RIS. The power of $\br_\RIS$ is
\begin{align}
\label{eq:RIS-tx-power}
P_{\RIS}(\bp,\bEtad) &\!=\! \EX{\norm{\bar{\br}_\RIS}^2} \nonumber \\
&\!=\! \sum\nolimits^K_{j=1} \etad_j \tr( \bP \bH\trans \bw_j \bw\herm_j \bH^{\ast} \bP\herm) \!+\! \sigma^2_\tR \tr(\bP\bP\herm) \nonumber \\
&\!=\! \bp\herm \boldsymbol{\Pi} \bp \,.
\end{align}
where
\begin{equation}
    \boldsymbol{\Pi} \!=\! \sum\nolimits^K_{j=1} \etad_j\, \diag(\bH\trans \bw_j) \left(\diag(\bH\trans\bw_j)\right)\herm + \sigma^2_\tR \bI_{\Nr}\,. \label{eq:PI}
\end{equation}
The available transmit power budget at the RIBS for the DL, $P_{\text{max}}$, has to be properly split between the BS and the active RIS to strike an effective balance between transmit and reflect precoding, respectively. Therefore, we next formulate the sum SE maximization problem with respect to $\bp$ and the fraction of the DL power budget allocated to the RIS, denoted by $\varepsilon$:%
\begin{subequations}%
\label{prob:P1:P:active}%
\begin{align}%
  \mathop {\max\limits_{\bp,\,\varepsilon}} & \quad \Sigma_{\mathsf{SE}}(\bp,\,\varepsilon) \!=\!\sum\nolimits^K_{k=1} \log_2 \Big(1 + \gamma_{k}(\bp,\,\varepsilon)\Big) \label{prob:P1:P:active:obj} \\
  \text{s.t.} 
    &\quad \bp\herm \boldsymbol{\Pi} \bp \leq \varepsilon P_{\text{max}} \, , \label{prob:P1:P:active:C1} \\
    &\quad 0\!<\varepsilon\!<1 \, .\label{prob:P1:P:active:C2}
\end{align}
\end{subequations}
Constraint~\eqref{prob:P1:P:active:C1} is convex in $\bp$ for a fixed choice of $\varepsilon$. Thus, problem~\eqref{prob:P1:P:active} can be efficiently solved by exploiting FP methods~\cite{Shen2018,ZhangZ2023}.
\setlength{\textfloatsep}{5mm}
\begin{algorithm}[t] \small
	\setstretch{1.06}
	\caption{RIS configuration and power split optimization}
	\vspace{1mm}
	\textbf{Input:} A randomly-drawn $\bp$, $\epsilon\!>\!0$, CSI to compute $\gamma_k(\bp)$, $\varepsilon_{\text{max}} \!=\! 1$ and $\varepsilon_{\text{min}} \!=\! 0$, $\mu_{\text{max}} \!=\! 10^4$ and $\mu_{\text{min}} \!=\! 0$, $P_{\text{max}}$, a precoding scheme.
	\begin{algorithmic}[1]		
		\Repeat {\texttt{~\%\%  Bisection over $\varepsilon$}}		
		\State $\varepsilon \gets (\varepsilon_{\text{min}} + \varepsilon_{\text{max}})/2$;
        \State Compute $\bw_{k}$, and $\etad_k$ as in~\eqref{eq:fractional-power-allocation}, $\forall k$;
        \State Compute $\gamma_k(\bp)$ as in~\eqref{eq:DL_SINR_PCSI}, $\forall k$;
        \State $\rho_k \gets \gamma_k,~\forall k$, and $\Sigma^{\mathsf{opt}}_{\mathsf{SE}}(\bp,\,\varepsilon) \gets \sum^K_{k=1} \log_2(1\!+\!\rho_k)$;
        \State Compute $\varpi_k$ and $\rho_k$ as in~\eqref{eq:optimal-varpi} and~\eqref{eq:optimal-rho}, respectively, $\forall k$;
        \Repeat {\texttt{~\%\%  Bisection over $\mu$}}
        \State $\mu \gets (\mu_{\text{min}} + \mu_{\text{max}})/2$;
        \State Compute $\bp^{\mathsf{opt}}$ as in~\eqref{eq:optimal-p} and $P_{\RIS}(\bp^{\mathsf{opt}},\bEtad)$ as in~\eqref{eq:RIS-tx-power};
        \If{$P_{\RIS}(\bp^{\mathsf{opt}},\bEtad) \!>\! \varepsilon P_{\text{max}}$}{~$\mu_{\text{min}} \gets \mu$;}
		\Else {~$\mu_{\text{max}} \gets \mu$};	
		\EndIf
        \Until $\mu_{\text{max}}\!-\!\mu_{\text{min}} \leq \epsilon$
        \State Update $\bw_{k}$, and $\etad_k$ as in~\eqref{eq:fractional-power-allocation}, $\forall k$;
        \State Update $\gamma_k(\bp^{\mathsf{opt}})$ as in~\eqref{eq:DL_SINR_PCSI}, $\forall k$;
        \State $\Sigma_{\mathsf{SE}}(\bp^{\mathsf{opt}},\,\varepsilon) \gets \sum^K_{k=1} \log_2(1\!+\!\gamma_k(\bp^{\mathsf{opt}}))$;
		\If{$\Sigma_{\mathsf{SE}}(\bp^{\mathsf{opt}},\,\varepsilon)\!<\!\Sigma^{\mathsf{opt}}_{\mathsf{SE}}(\bp,\,\varepsilon)$}{~$\varepsilon_{\text{max}} \gets \varepsilon$;}
		\Else		
		\State $\varepsilon_{\text{min}} \gets \varepsilon$ and $\bp \gets  \bp^{\mathsf{opt}}$;
		\State $\Sigma^{\mathsf{opt}}_{\mathsf{SE}}(\bp,\,\varepsilon) \gets \Sigma_{\mathsf{SE}}(\bp^{\mathsf{opt}},\,\varepsilon)$;		
		\EndIf
		\Until $\varepsilon_{\text{max}}\!-\!\varepsilon_{\text{min}} \leq \epsilon$
	\end{algorithmic}
	\textbf{Output:} $\bp^{\mathsf{opt}}$, $\varepsilon$.
	\label{alg:RIS_opt_active}
\end{algorithm}
By introducing auxiliary variables $\boldsymbol{\rho} \!=\! [\rho_1,\ldots,\rho_k] \!\in\! \mR^K_{+}$ and $\boldsymbol{\varpi} \!=\! [\varpi_1,\ldots,\varpi_K] \!\in\! \mC^K$, and using the Lagrangian dual reformulation to reshape the objective function, problem~\eqref{prob:P1:P:active} can be equivalently formulated as
\begin{subequations} \label{prob:P2:P:active}
\begin{align}	
  \mathop {\max\limits_{\substack{\bp,\,\varepsilon\\\boldsymbol{\rho},\,\boldsymbol{\varpi}}}} & \quad 
  \sum\limits^K_{k=1} \ln(1 \!+\! \rho_k) \!-\! \sum\limits^K_{k=1} \rho_k \!+\! \sum\limits^K_{k=1} g(\bp,\rho_k,\varpi_k)\label{prob:P2:P:active:obj} \\
  \text{s.t.}
    &\quad \bp\herm \boldsymbol{\Pi} \bp \leq \varepsilon P_{\text{max}} \, , \label{prob:P2:P:active:C1} \\
    &\quad 0\!<\varepsilon\!<1 \, ,\label{prob:P2:P:active:C2}
\end{align}
\end{subequations}
where $g(\bp,\rho_k,\varpi_k) \!=\! 2\sqrt{1\!+\!\rho_k}\, \xi_k \!-\! |\varpi_k|^2 \mathcal{I}_k(\bp)$,
and we have defined $\xi_k \!=\! \sqrt{\etad_k}\,\re{\varpi_k^{\ast}\bp\trans\bG\trans_k\bw_k}$, and
\begin{equation}
    \mathcal{I}_k(\bp) \!=\! \sum\limits_{j=1}^K \etad_j\left|\bp\trans	\bG_k\trans \bw_j \right|^2 \!+\! \norm{\bp\trans \bH_k}^2\! \sigma^2_\tR \!+\! \sigma^2_k\,.
\end{equation}
For a fixed value of $\varepsilon$, $\boldsymbol{\rho}$ and of $\bp$, the optimal $\boldsymbol{\varpi}$ can be obtained by equating to zero the partial derivative of~\eqref{prob:P2:P:active:obj} with respect to $\varpi_k$, $k\!=\!1,\ldots,K$, as
\begin{align}
    \label{eq:optimal-varpi}
    \varpi^{\mathsf{opt}}_k &\!=\! \sqrt{\etad_k(1\!+\!\rho_k)}\,\frac{\bp\trans\bG\trans_k\bw_k}{\mathcal{I}_k(\bp)}\,,\; k\!=\!1,\ldots,K.
\end{align}
For a fixed value of $\varepsilon$, $\bp$ and of $\boldsymbol{\varpi}$, the optimal $\boldsymbol{\rho}$ can be obtained by equating to zero the partial derivative of~\eqref{prob:P2:P:active:obj} with respect to $\rho_k$, $k\!=\!1,\ldots,K$, as
\begin{align}
    \label{eq:optimal-rho}
    \rho^{\mathsf{opt}}_k \!=\! \frac{1}{2}\xi^2_k \!+\!\frac{1}{2}\xi_k\sqrt{\xi^2_k \!+\!4},\; k\!=\!1,\ldots,K.
\end{align}
For a fixed value of $\varepsilon$, $\boldsymbol{\rho}$ and $\boldsymbol{\varpi}$, the sum SE maximization problem can be reformulated, with respect to $\bp$, as
\begin{subequations} \label{prob:P3:P:active}
\begin{align}	
  \mathop {\max\limits_{\bp}} & \quad \re{2\bp\herm\boldsymbol{\upsilon}} \!-\! \bp\herm \bOmega \bp\label{prob:P3:P:active:obj} \\
  \text{s.t.}
    &\quad \bp\herm \boldsymbol{\Pi} \bp \leq \varepsilon P_{\text{max}} \, , \label{prob:P3:P:active:C1}
\end{align}
\end{subequations}
where
\begin{align}
    \boldsymbol{\upsilon} &\!=\! \sum\nolimits^K_{k=1} \sqrt{\etad_k(1\!+\!\rho_k)}\,\varpi_k^{\ast} \bG\trans_k\bw_k\,, \label{eq:upsilon} \\
    \bOmega &\!=\! \sum^K_{k=1} |\varpi_k|^2 \bH\herm_k \bH_k \sigma^2_\tR \!+\! \sum^K_{k=1} |\varpi_k|^2 \bG\trans_k \bar{\bW} \bG^{\ast}_k\,, \label{eq:Omega} 
\end{align}
and $\bar{\bW}\!=\! \sum\nolimits^K_{j=1} \etad_j \bw_j\bw\herm_j$.
The above problem is a \textit{quadratically constrained quadratic program}, whose optimal solution can be obtained by adopting the Lagrange multiplier method and in closed form as~\cite{ZhangZ2023} 
\begin{equation}
    \label{eq:optimal-p}
    \bp^{\mathsf{opt}} \!=\! (\bOmega \!+\!\mu\boldsymbol{\Pi})^{-1}\boldsymbol{\upsilon}\, ,
\end{equation}
where $\mu$ is the Lagrange multiplier and its optimal value can be obtained via a binary search such that the complementary slackness condition of constraint~\eqref{prob:P3:P:active:C1} is met.

\subsection{Heuristic DL Power Allocation}
\label{subsec:power-control}
Transmit precoding at the BS and reflect precoding at the RIS are closely entangled as they are under a cumulative power constraint. Indeed, $\bp$ and $\bEtad$ are coupled in the RIS power constraint~\eqref{prob:P1:P:active:C1}. Moreover, $\bEtad$ must also satisfy the following power constraint at the BS:
\begin{align}	
\sum\nolimits^K_{k=1} \etad_k \!\leq\! (1\!-\!\varepsilon)P_{\text{max}} \,. 
\label{eq:BS-power-constraint}
\end{align}
A heuristic power allocation strategy consists in setting
\begin{equation}
    \label{eq:fractional-power-allocation}
    \etad_k \!=\! \min\left\{(1\!-\!\varepsilon)P_{\text{max}} \varrho_k(\bp), \frac{\varepsilon P_{\text{max}} \!-\! \norm{\bp}^2\sigma^2_{\tR}}{\norm{(\bH\bP)\trans \bw_k}^2} \varrho_k(\bp)  \right\},\!
\end{equation}
where we have defined
\begin{equation}
    \varrho_k(\bp) \!=\! \frac{\left[\tr\left(\bar{\bh}_k\bar{\bh}_k\herm\right)\right]^{\nu}}{\sum\nolimits^K_{i=1} \left[\tr\left(\bar{\bh}_i\bar{\bh}_i\herm\right)\right]^{\nu}}\,,
\end{equation}
with the parameter $\nu$ establishing the power allocation policy. If $\nu > 0$, the BS allocates more power to the UEs with better channel gains. Such a choice would be in line with the sum SE maximization objective. Importantly,~\eqref{eq:fractional-power-allocation} guarantees that the constraints~\eqref{prob:P1:P:active:C1} and~\eqref{eq:BS-power-constraint} are simultaneously satisfied.

\subsection{Joint Optimization of RIS Configuration and Power Split}
Algorithm~\ref{alg:RIS_opt_active} summarizes the steps taken to jointly optimize the configuration of the active RIS and the power split factor, $\varepsilon$. Initially, $\bp$ is randomly drawn. While the optimal value of $\varepsilon$ is found via a bisection search (i.e., the outer loop of Algorithm~\ref{alg:RIS_opt_active}), in each iteration of this search, for a fixed value of $\varepsilon$, the optimal RIS configuration is obtained according to~\eqref{eq:optimal-p}. The latter is, in turn, iteratively refined with respect to $\mu$ via binary search (i.e., the inner loop of Algorithm~\ref{alg:RIS_opt_active}). Notably, strong convergence of the FP methods has been proved in~\cite{Shen2018}, hence Algorithm~\ref{alg:RIS_opt_active} eventually converges. If the updates in each outer-loop iteration step of $\boldsymbol{\varpi}$, $\boldsymbol{\rho}$ and $\bp$ are optimal, a local optimum to problem~\eqref{prob:P2:P:active} is obtained. Unlike~\cite{ZhangZ2023}, which jointly optimizes the reflect and transmit precoding vectors, we adopt heuristic schemes for the latter to keep the complexity of the proposed algorithm practical. 

\section{Numerical Results}
\label{numerical-results}

\subsection{Simulation Scenario and Settings}
Unless stated otherwise, we consider a BS with $\Na\!=\!16$ antennas and an active RIS with $\Nr\!=\!64$ elements serving $K\!=\!25$ outdoor UEs, all with isotropic antennas. The BS–RIBS distance $D$ ranges from $4\lambda$ to $5\lambda$, based on the RIS aperture~\cite{interdonato2024RISaided}, and the BS tilt angle is $\alpha\!=\!\pi/6$. The RIBS is 30 m high at $(-158,\, 8)$ m, and UEs are 1.5 m high, randomly placed within $(-90,\,-160)$ to $(150,\, 210)$ m, ensuring a minimum 3D RIBS–UE distance of 70 m. The same deployment applies to both stochastic and ray-tracing channel models.
For the stochastic model, ${\beta_k}$ follow the urban model in\cite{3GPP_25996_model}, which also defines LoS probability and Rician factor. The standard deviations of azimuth and elevation angles are $15^{\circ}$, implying strong spatial correlation, and $S_k\!=\!1$ for all the UEs.
The bandwidth is 1 MHz, and the noise power is $-107$ dBm at both RIS and UEs. Lastly, $P_{\text{max}}\!=\!0.5$ W and $\nu\!=\!0.5$.

\subsection{Performance Evaluation}

This section presents the performance evaluation of the RIBS under both statistical and ray-tracing channel models, emphasizing the gains from RIBS optimization and the realism of ray-tracing evaluations.

\begin{figure} 
    \centering
  \subfloat[Rician-fading channel model.\label{fig:precoding-Rician}]{%
       \includegraphics[width=.8\linewidth]{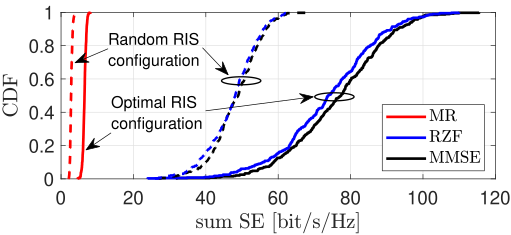}}\\
    \vspace{-2mm}
  \subfloat[Ray-Tracing channel model.\label{fig:precoding-Sionna}]{%
        \includegraphics[width=.8\linewidth]{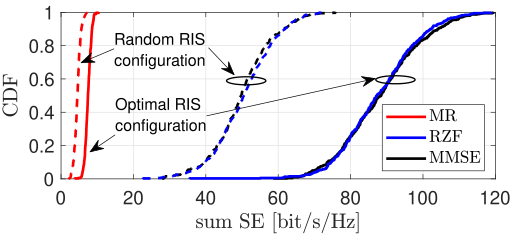}}
  \caption{CDFs of the sum SE for different precoders and RIS configurations.}
  \label{fig1:sumSE-CDFs}
  \vspace{-2mm}
\end{figure}

Fig.~\ref{fig1:sumSE-CDFs} compares the cumulative distribution functions (CDFs) of the DL sum SE across multiple UE deployments, considering both random and optimized RIS configurations under the statistical and SIONNA ray-tracing models. The ``random RIS configuration'' refers to a randomly drawn $\bp$ with $\varepsilon\!=\!0.25$, while ``optimal RIS configuration'' corresponds to the output of Algorithm~\ref{alg:RIS_opt_active}. Three precoding schemes are considered: maximum ratio (MR), regularized zero-forcing (RZF), and minimum mean-square error (MMSE).
%
%
As shown in Figs.~\ref{fig:precoding-Rician}--\ref{fig:precoding-Sionna}, optimized RIS configurations consistently outperform random ones, with a median sum SE gain over 50\%. While the statistical model yields modest SE due to its conservative assumptions, the ray-tracing model reveals significantly higher performance, highlighting the importance of environment-aware design. The median SE nearly doubles under ray tracing with optimal RIS, highlighting the importance of proper RIBS configuration and the efficiency of ray-tracing-based optimization.
Lastly, it is observed, although the results are omitted due to space constraints, that the optimal value of $\varepsilon$, as output by Algorithm~\ref{alg:RIS_opt_active}, is consistently higher than 0.5 across all RIS configurations, precoders, and channel models. This result demonstrates that allocating a larger fraction of the power budget to the RIS is beneficial, as the active RIS plays a dominant role in suppressing interference, thereby enhancing the SE.

\begin{figure} 
    \centering
  \subfloat[sum SE versus $\Nr$. Here, $\Na = 16$.\label{fig:Nr}]{%
       \includegraphics[width=.785\linewidth]{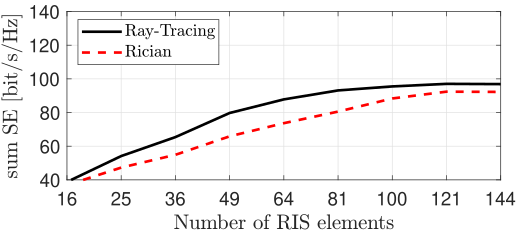}}\\
    \vspace{-1mm}
  \subfloat[sum SE versus $P_{\text{max}}$. Here, $\Na = 16$ and $\Nr = 64$.\label{fig:Pmax}]{%
        \includegraphics[width=.785\linewidth]{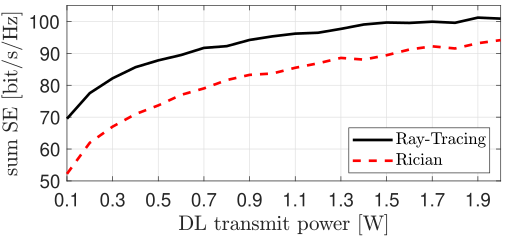}}
  \caption{Optimal average sum SE for different channel models.}
  \label{fig2:sumSE-AVGs} 
  \vspace{-2mm}
\end{figure}
Fig.~\ref{fig2:sumSE-AVGs} analyzes the impact of two parameters: the number of RIS elements $\Nr$ and the total DL power budget $P_{\text{max}}$, using RZF precoding and optimized RIBS configuration. In Fig.\ref{fig:Nr}, the average sum SE increases with $\Nr$ due to improved beamforming gain, but eventually saturates as interference mitigation becomes less effective under stronger spatial correlation. Also, the SEs under both models converge as $\Nr$ grows.
Fig.~\ref{fig:Pmax} shows that increasing $P_{\text{max}}$ improves SE for both models, but the gains diminish at high power levels due to greater interference. Again, the performance under both models converges at large $P_{\text{max}}$ values.

\section{Conclusion}
\label{conclusion}
This paper evaluated the performance of an RIBS, an architecture that reduces RF chains while maintaining mMIMO performance. Using the SIONNA ray tracing module, we compared RIBS's performance against 3GPP-compliant statistical models. In the simulated scenario, ray tracing predicted higher performance than 3GPP-compliant statistical models, a result that still awaits confirmation through empirical measurements. Ongoing work focuses on RIBS energy savings compared to conventional mMIMO and developing site-specific transceiver algorithms.

\linespread{0.98}
\bibliographystyle{IEEEtran}
\bibliography{IEEEabrv,main}

\end{document}